\documentclass{article}
\usepackage{amsmath}

\renewcommand{\a}{\alpha} 
\renewcommand{\b}{\beta}
\newcommand{\g}{\gamma} 
\newcommand{\G}{\Gamma}
\renewcommand{\d}{\delta} 

\renewcommand{\k}{\kappa} 
\renewcommand{\l}{\lambda} 
 
\newcommand{\m}{\mu} 
\newcommand{\n}{\nu}
\newcommand{\p}{\pi} 
 
\newcommand{\s}{\sigma}
\renewcommand{\S}{\Sigma} 

\renewcommand{\t}{\tau}

\newcommand{\cL}{{\cal L}}

\newcommand{\pd}{\partial}

\newcommand{\kummer}{\mbox{}_1\mbox{F}_1}
\newcommand{\half}{\frac{1}{2}}
\newcommand{\shalf}{{\textstyle\frac{1}{2}}}

\newcommand{\bS}{\bar{S}}

\begin{document}

\author{Jiri Hoogland\footnote{jiri@cwi.nl} 
  and Dimitri Neumann\footnote{neumann@cwi.nl}\\
  CWI, P.O.~Box 94079, 1090 GB  Amsterdam, The Netherlands}
\title{\textbf{Asians and cash dividends:\\Exploiting symmetries
in pricing theory}}
\maketitle

\thispagestyle{empty}
\begin{abstract}
In this article we present new results for the pricing
of arithmetic Asian options within a Black-Scholes context.
To derive these results we make extensive use of the local
scale invariance that exists in the theory of contingent
claim pricing. This allows us to derive, in a natural way,
a simple PDE for the price of arithmetic Asians options. 
In the case of European average strike options, a proper
choice of numeraire reduces the dimension of this PDE to
one, leading to a PDE similar to the one derived by Rogers
and Shi. We solve this PDE, finding a Laplace-transform
representation for the price of average strike options,
both seasoned and unseasoned. This extends the results of
Geman and Yor, who discussed the case of average price options. 
Next we use symmetry arguments to show that prices of average
strike and average price options can be expressed in terms
of each other. Finally we show, again using symmetries,
that plain vanilla options on stocks paying known cash
dividends are closely related to arithmetic Asians, so that
all the new techniques can be directly applied to this case.
\end{abstract}

\newpage

\section{Introduction.}

Options with a payoff-function which depends on the average of some
underlying, a.k.a. Asian-type option have a multitude of applications
in finance. They find applications for example in currency-based
contracts, interest rates and commodities. In the following we will
consider a setting where stock prices are modeled by geometric
Brownian motions. Depending on the type of averaging the analytic
price of such a contract is easy or difficult to compute. Geometric
averaging leads to simple expressions for the
prices~\cite{HooglandNeumann99b}. Arithmetic averaging however is a
highly non-trivial exercise and one has to rely on either
approximations~\cite{TurnbullWakeman91}, partially
analytic~\cite{GemanYor93} or numerical
solutions~\cite{Curran94,RogersShi95}. In Ref.~\cite{RogersShi95} a
simple PDE was derived and bounds on prices of an average price option
were derived. These bounds have been improved in
Ref.~\cite{Thompson99}.

\vspace{1\baselineskip}\noindent
In this article we provide an alternative approach to derive
(partially) analytic solutions of Asian-type contracts with arithmetic
averaging. Using the fundamental notion of a local scale
invariance~\cite{HooglandNeumann99a,HooglandNeumann99b} we derive a
general PDE for a (European) Asian-type contract with arithmetic
averaging. This result is then linked to the PDE derived by Rogers and
Shi~\cite{RogersShi95} and we proceed by solving the Laplace-transform
of the solution of this PDE for the case of an average strike put,
both for the unseasoned and seasoned case. This extends the results of
Geman and Yor, which gave the solution for average price options.
Next we show that the local scale invariance allows one to identify
the average strike call and average price put by substitution of the
proper parameters. This is a new result. Finally we consider the
problem of a vanilla option on a stock which pays known cash
dividends. Again the local scale symmetry allows one to relate the
value of such a contract to that of arithmetic average options. 

\section{Homogeneity and contingent claim pricing}
\label{sec:homogeneity}

In previous papers~\cite{HooglandNeumann99a,HooglandNeumann99b}
we have shown that a fundamental property of
any properly defined market of tradables\footnote{
  Tradables are objects which are trivially self-financing:
  it doesn't cost nor yield money to keep a fixed amount of them.
  Examples are stocks and bonds. Note that money is not a tradable,
  unless the interest rate is zero.} 
is that the price of any claim depending on other tradables in
the market should be a homogeneous\footnote{
  A function $f(x_0,\ldots,x_n)$ is called
  homogeneous of degree $r$ if $f(ax_0,\ldots,ax_n)=a^r
  f(x_0,\ldots,x_n)$. Homogeneous functions of degree $r$ satisfy the
  following property (Euler): $\sum_{\m=0}^nx_\m\frac{\pd}{\pd
  x_\m}f(x_0,\ldots,x_n)=r f(x_0,\ldots,x_n)$}
function of degree one
of these same tradables. This property is nothing but a consequence
of the simple fact that prices of tradables are only defined
with respect to each other. Let us review some of the content
of Ref.~\cite{HooglandNeumann99a}. Assume that we have a market of
$n+1$ basic tradables with prices $x_\mu$ ($\mu=0,\ldots,n$) at time
$t$. The price of any tradable in this market with a payoff depending
on the prices of these basic tradables should satisfy the following
scaling symmetry:
\[
V(\l x,t)=\l V(x,t)
\]
which automatically implies\footnote{We make use of Einsteins
  summation convention: repeated indices in products are
  implicitly summed over, unless stated otherwise.} (Euler)
\[
V(x,t)=x_\m \pd_{x_\m}V(x,t)
\]
where $\pd/\pd x_\m\equiv\pd_{x_\m}$. This is a universal
property, independent of the choice of dynamics. We use this fundamental
property to derive a general PDE, giving the price of such a claim
in a world where the dynamics of the tradables are driven
by $k$ independent standard Brownian motions, as
follows\footnote{Both the $\s_\m$ and $dW$ are vectors, the dot
denotes an inner-product w.r.t. the $k$ driving diffusions.}
\[
dx_\m(t)=x_\m(t)\big(\s_\m(x,t)\cdot dW(t)+\a_\m(x,t) dt\big),
\hspace{5mm}\mbox{(no sum)}
\]
Consistency requires that both $\s_\m$ and $\a_\m$ are
homogeneous functions of degree zero in the tradables, i.e.
they should only depend on ratios of prices of tradables.
Note that we do not specify the numeraire in terms of which
the drift and volatility are expressed. This choice is
irrelevant for the pricing problem, as we will see.
Applying It\^o to $V(x,t)$ we get
\[
dV(x,t)=\pd_{x_\m}V(x,t)dx_\m+\cL V(x,t)dt
\]
where 
\[
\cL V(x,t)\equiv 
\bigg(
\pd_t+\half\s_\m(x,t)\cdot\s_\n(x,t)x_\m x_\n\pd_{x_\m}\pd_{x_\n}
\bigg)V
\]
So, if $V(x,t)$ solves $\cL V=0$ with the payoff at maturity as
boundary condition $V(x,T)=f(x)$, then we immediately have a
replicating self-financing trading strategy because of the homogeneity
property. We will drop the distinction between such derived and basic
quantities and always refer to them as tradables. Note that we do not
have to use any change of measure to arrive at this result, by keeping
the symmetry explicit. Drifts are irrelevant for the derivation of the
claim price. Only the requirement of uniqueness of the solution,
i.e. no arbitrage, leads to constraints on the drifts terms if
deterministic relations exist between the various
tradables~\cite{HooglandNeumann99a}.

\subsection{Symmetries of the PDE}
\label{sec:symmetries-pde}
The scale invariance of the claim price is inherited by the PDE
via an invariance of the solutions of the PDE under a simultaneous
shift of all volatility-functions by an arbitrary function $\l(x,t)$
\begin{equation}
\s_\m(x,t)\to\s_\m(x,t)-\l(x,t) \label{eq:sbst}
\end{equation}
Indeed, if $V$ is solves $\cL V=0$, then it also solves
\[
\bigg(
\pd_t+\half(\s_\m(x,t)-\l(x,t))\cdot(\s_\n(x,t)-\l(x,t))
x_\m x_\n\pd_{x_\m}\pd_{x_\n} \bigg)V=0
\]
This can easily be checked by noting that for homogeneous
functions of degree~1 we have
\[
x_\m \pd_{x_\m}\pd_{x_\n}V=0
\]
This ensures that terms involving the $\l$ drop out of the PDE. (Note
that this equation gives interesting relations between the various
$\Gamma$'s of the claim). From this it follows that $V$ itself must be
invariant under the substitution defined by Eq.~\ref{eq:sbst}.
This corresponds to the freedom of choice of a numeraire. It
just states that volatility is a relative concept. Price functions
should not depend on the choice of a numeraire.

\subsection{The algorithm}
\label{sec:algorithm}

To price contingent claims we start out with a basic set of
tradables. Using these tradables we may construct new, derived,
tradables, whose price-process $V$ depends upon the basic tradables.
Of course, these new tradables should be solutions to the basic PDE,
$\cL V=0$. Their payoff functions serve as boundary conditions.
(Note that prices of basic tradables trivially satisfy the PDE,
by construction). If the derived tradables are constructed in this
way, we can use them just like any other tradable. In particular,
we can use them as underlying tradables, in terms of which the price
of yet other derivative claims can be expressed (and so on...)
In fact, this is a fundamental property that any correctly defined
market should posses. It amounts to a proper choice of
coordinates to describe the economy.

\vspace{1\baselineskip} \noindent
The general approach to the pricing of a path-dependent
claim in our formalism can be described as follows.
\begin{enumerate}
\item The payoff is written in terms of tradable objects.
\item A PDE is derived for the claim price with respect to these tradables. 
\item The PDE is solved.
\item Possible consistency check: the solution should be invariant
under the substitution Eq.~\ref{eq:sbst} (numeraire independence).
\end{enumerate}

\subsection{Generalized put-call symmetries}
\label{sec:putcall}

As an example of the strength of this symmetry, and to show the natural
embedding in our formalism, consider an economy with two tradables
with prices denoted by $x_{1,2}$ and dynamics given by ($i=1,2$)
\[
dx_i(t)=x_i(t)\s_i(x_1,x_2,t) \cdot dW(t)+\ldots
\hspace{5mm}\mbox{(no sum)}
\]
It is easy to see that under certain conditions there should be a
generalized put-call symmetry. Any claim with payoff $f(x_1,x_2)$ at
maturity and price $V(x_1,x_2,t)$ should satisfy
\[
\bigg(\pd_t+\half |\s(x_1,x_2,t)|^2 x_1^2 \pd_{x_1}^2\bigg)V=0
\]
where $\s(x_1,x_2,t)\equiv\s_1(x_1,x_2,t)-\s_2(x_1,x_2,t)$. 
Homogeneity implies that it also solves
\[
\bigg(\pd_t+\half |\s(x_1,x_2,t)|^2 x_2^2 \pd_{x_2}^2\bigg)V=0
\]
Therefore, if $|\s(x_1,x_2,t)|^2=|\s(x_2,x_1,t)|^2$, this PDE can be
rewritten as
\[
\bigg(\pd_t+\half |\s(x_2,x_1,t)|^2 x_2^2 \pd_{x_2}^2\bigg)V=0
\]
and we see that $V(x_2,x_1,t)$ with payoff $f(x_2,x_1)$ is a
solution, too. This is nothing but a generalized put-call symmetry.
In the first case $x_2$ acts as numeraire, in the second case $x_1$
takes over this role. The usual put-call symmetry follows if we
take a constant $\s$ and let $x_1, x_2$ represent a stock and a bond
respectively. 

\subsection{Lognormal asset prices}
\label{sec:lap}
In an economy with lognormal distributed asset-prices
\[
dx_\m(t)=x_\m(t)\s_\m(t)\cdot dW(t)+\ldots
\hspace{5mm}\mbox{(no sum)}
\]
it is possible to write down a very elegant formula for European-type
claims, as was shown in Ref.~\cite{HooglandNeumann99a}
\begin{equation}
  \label{eq:1}
  V(x_0,\ldots,x_n,t)=\int V(x_0 \phi(z-\theta_0),\ldots,
  x_n \phi(z-\theta_n),T)d^mz
\end{equation}
with
\[
\phi(z)=\frac{1}{\left(\sqrt{2\pi}\right)^m}
\exp\left( -\frac{1}{2} \sum_{i=1}^{m} z^2_i \right)
\]
The $\theta_\m$ are $m$-dimensional vectors, which follow from a
singular value decomposition of the covariance matrix $\S_{\m\n}$ of
rank $m\le k$:
\[
\S_{\m\n}\equiv\int_t^T \s_{\m}(u)\cdot\s_{\n}(u)du=\theta_\m\cdot\theta_\n
\]

\section{Arithmetic Asians}

In this section we will consider the pricing of
Arithmetic Asian options. Since this is the only type
of Asian options that we will look at, we will omit
the word 'Arithmetic' in the sequel. Note that parts of this
material already appeared in Ref.~\cite{HooglandNeumann99b}.
A fundamental building block in the construction of a European
Asian option, expiring at time $T$, is a tradable which at
time $T$ represents the value of a stock at an earlier time $s$.
But to define this, we must first agree how to translate value
through time. For this, we need a reference asset. A convenient
choice is to take a bond $P(t,T)$ (or $P(t)$ for short), which
matures at time $T$, as reference. Then we can define
$$
  Y_s(t) = \left\{ \begin{array}{ll} S(t) & t<s \\
  \frac{S(s)}{P(s)} P(t) & t\geq s \end{array} \right.
$$
In words, this is a portfolio where one starts out with a
stock, and converts it into a bond at time $t=s$. It is
trivially self-financing. To set the stage, we will assume
that the interest rate has constant value $r$, as is usual
in the Black-Scholes context (stochastic interest rates are
much harder to handle, see Ref.~\cite{HooglandNeumann99b}).
In that case the bond with maturity $T$ has value
$e^{-r(T-t)}$ when expressed in the currency in which it is
nominated (say dollars). Consequently, an amount
$e^{-r(T-s)}$ of the tradable $Y_s$
will have a dollar value at time $T$ which is equal to the
dollar value of the stock at time $s$. We will also assume
that the contracts are initiated at time $t=0$, unless
stated otherwise. Note that for $t\in [0,T]$ we have
\begin{equation}\begin{split}
  S(t) &= Y_T(t) \\
  P(t) &= \frac{P(0)}{S(0)} Y_0(t)
\end{split} \label{eq:yrs} \end{equation}

\subsection{Discretely sampled Asians}

The main reason for introducing the objects $Y_s(t)$
is that they constitute a natural basis of tradables
in which prices and payoffs of arithmetic Asians can be
expressed. For example, the payoff of a discretely
sampled average price call (APC) can be written as
$$
 \left(\sum_i w(t_i) Y_{t_i}(T)-KP(T) \right)^+
=\left(\left(\sum_i w(t_i) \frac{S(t_i)}{P(t_i)}-K\right)
  P(T)\right)^+
$$
where $\{t_i\}$ is a set of sample times, $w(t_i)$ are
corresponding weights, and $K$ is the strike. We use the
notation $(\cdot)^+$ for $\max(\cdot,0)$. Observe that the
payoff explicitly contains $P(T)=1\$$ to make it homogeneous
of degree one in the tradables. In a similar way, the payoff
of an average strike put (ASP) becomes
$$
 \left(\sum_i w(t_i) Y_{t_i}(T)-kS(T) \right)^+
$$
where $k$ is a (generalized) strike. In view of Eqs.~\ref{eq:yrs},
we see that both options are in fact instances of a more general
discrete Asian option, which is defined by the payoff
\begin{equation}
 V(\{t_i\},w,T)=
   \left(\sum_i w(t_i) Y_{t_i}(T) \right)^+
  \label{eq:genasia}
\end{equation}
 
\subsection{Valuation by multiple integrals}
\label{sec:mulint}

In this section we want to calculate the value of the
generalized option defined by Eq.~\ref{eq:genasia}, at the
time the contract is initiated, i.e. $t=0$. This value
is known as its {\it unseasoned} value. Without loss
of generality, we will assume that there are $N+1$
sample times, satisfying $0=t_0<t_1<\cdots<t_{N-1}<t_N=T$.
If we take the bond as numeraire, and assume
that the stock price follows a lognormal price process
$$
  dS(t) = \s S(t) dW(t) + \cdots
$$
then it is easy to derive that the tradables $Y_s(t)$ satisfy
$$
  dY_s(t) = {\bf 1}_{t<s}\s Y_s(t) dW(t) + \cdots
$$
where ${\bf 1}_{t<s}$ is the indicator function. It equals
one when $t<s$ and zero otherwise. So the $Y_s(t)$ also follow
lognormal price processes with a time dependent volatility,
and we can directly use the results of section~\ref{sec:lap}.
The first step is to calculate the variance-covariance matrix
of the tradables
$$
  \S_{ij} = \s^2 \int_0^T {\bf 1}_{t<t_i}{\bf 1}_{t<t_j} dt
  = \s^2 \min(t_i,t_j)
$$
The rank of this matrix is $N$. A singular value decomposition
$\S_{ij}=\theta_i\cdot\theta_j$ is given by
$$
  \theta_0 = ( \underset{N}{\underbrace{0,\ldots,0}} ), \hspace{5mm}
  \theta_i = ( \b_1,\ldots,\b_i,
               \underset{N-i}{\underbrace{0,\ldots,0}} )
$$
and the $\b_i$ are defined by
$$
  \b_i = \s\sqrt{t_i-t_{i-1}}
$$
The value of the unseasoned option can now be written as an
$N$-dimensional integral (here $z=(z_1,\ldots,z_N)$)
\begin{equation}
  V(\{t_i\},w,0) =
  S(0) \int d^Nz
  \left( \sum_{i=0}^N w(t_i) \phi(z-\theta_i) \right)^+ 
  \label{eq:unseas}
\end{equation}
where we used the fact that $Y_s(0)=S(0)$ for all $0\leq s\leq T$.
This is in fact a Feynman-Kac formula.

\subsection{An interesting duality}
\label{sec:tdual}

In this section, we will again calculate the unseasoned value of
the option in Eq.~\ref{eq:genasia}, but this time we take the
stock as numeraire. This corresponds to a shift of $\s$ in all
volatility functions. We find that $Y_s(t)$ now satisfies
$$
  dY_s(t) = {\bf 1}_{t>s}\s Y_s(t) dW(t) + \cdots
$$
In this case the variance-covariance matrix becomes
$$
  \hat{\S}_{ij} = \s^2 \int_0^T {\bf 1}_{t>t_i}{\bf 1}_{t>t_j} dt
  = \s^2 \min(T-t_i,T-t_j)
$$
and a singular value decomposition
$\hat{\S}_{ij}=\hat{\theta}_i\cdot\hat{\theta}_j$ is given by
$$
  \hat{\theta}_i = ( \underset{i}{\underbrace{0,\ldots,0}},
  \b_{i+1},\ldots,\b_N ), \hspace{5mm}
  \hat{\theta}_N = ( \underset{N}{\underbrace{0,\ldots,0}} )
$$
So an alternative expression for the price of the option is
$$
  V(\{t_i\},w,0) =
  S(0) \int d^Nz
  \left( \sum_{i=0}^N w(t_i) \phi(z-\hat{\theta}_i) \right)^+ 
$$
However, if we compare this result with Eq.~\ref{eq:unseas},
we see that it could also be interpreted as the value of an
option with payoff
$$
  \left(\sum_{i=0}^{N} w(t_i) Y_{T-t_i}(T) \right)^+
$$
In other words, two options which are related by the
following substitution in their payoff
\begin{equation}
  \boxed{ Y_t(T) \leftrightarrow Y_{T-t}(T) } \label{eq:dual}
\end{equation}
have the same value at $t=0$. We will call this
T-duality (T from Time-reversal). It is a very interesting
symmetry operation because, in view of Eqs.~\ref{eq:yrs},
we can use it to relate the values of unseasoned
average strike and average price options. We will come back
to this point in section~\ref{sec:tdual2}. Note that
Eq.~\ref{eq:dual} takes a simple form by virtue of the fact
that we are working in a basis of tradables.

\subsection{A PDE approach}
\label{sec:pdeapp}

In this section we consider a PDE approach to the
pricing of Asian options. We derive a very
general PDE, which can be related to the one that is
usually found in the literature. Our PDE, however, has
the advantage of being manifestly numeraire independent
by virtue of the fact that it is expressed in a basis
of tradables. It can be used to price both American
and European style options, but we will focus on the
European case here. We will come back to American
Asians in future work. The basic idea in the derivation
of the PDE is, instead of introducing a tradable for
each sample date, to introduce one new tradable
$\bS(t)$, which is a weighted sum over $Y_s(t)$
(obviously, a sum of tradables is again a tradable).
This allows us to consider continuously sampled Asians
in a proper way. Also
$$
  \bS(t) = \int_0^T w(s) Y_s(t) ds \equiv
  \phi(t) S(t) + A(t) P(t)
$$
where $A(t)$ is proportional to the running average
$$
  A(t) = \int_0^t w(s) \frac{S(s)}{P(s)} ds
$$
and
$$
  \phi(t)=\int_t^T w(s) ds 
  \hspace{5mm}\longleftrightarrow\hspace{5mm}
  w(t) = -\partial_t \phi(t), \hspace{3mm} \phi(T)=0
$$
Of course this approach also incorporates discretely sampled
Asians. In that case, $w(t)$ will be a sum of Dirac
delta-functions and $\phi(t)$ will be a piecewise constant
function, making jumps at sample dates. If we choose the bond
$P(t)$ as numeraire, and assume that $S(t)$ satisfies
$$
  dS(t) = \sigma S(t) dW(t) + \cdots
$$
then it is obvious that the new tradable $\bS(t)$ satisfies
$$
  d\bS(t) = \phi(t) \sigma S(t) dW(t) + \cdots
$$
This straightforwardly leads to the following PDE for
options which depend on $\bS,S$ and $P$
\begin{equation}
\left(\partial_t+\half\s^2S^2
(\partial_S+\phi\partial_{\bS})^2\right)V=0 \label{eq:funpde}
\end{equation}
If we perform a change of variables in the PDE, eliminating
$\bar{S}$ in favor of $A$, we find
$$
\left(\partial_t+w\frac{S}{P}\partial_A+
\half\s^2S^2\partial^2_S\right)V=0
$$
which is closer to the usual formulation. However,
since $A(t)$ does not correspond to the value of a tradable
object, this form of the PDE looses the manifest symmetry.

\subsection{Analytical solutions}

In this section we derive a Laplace transform representation
for prices of average strike options in the case that sampling is
continuous with an exponential weight function. This is in fact
the counterpart of the calculation of Geman and Yor~\cite{GemanYor93}
for the price of average price options, although they used
entirely different methods to derive it. The results for
average strike options turn out to be somewhat more involved, 
mainly because there is no simple relation between prices
of seasoned and unseasoned options, while a simple relation
does exist in the case of average price options, as we will see.
Now, by definition, the payoff of an average strike option is
defined in terms of $\bS$ and $S$ only, $P$ does not appear. For
example, the payoff of an ASP is given by
\begin{equation}
  (\bS(T)-kS(T))^+ \label{eq:payoff}
\end{equation}
Because of this fact, it is natural to choose $S$ as numeraire.
Dropping the derivative w.r.t. $P$, the PDE then reduces to
$$
  \left(\partial_t+\half\s^2(\bS-\phi S)^2
  \partial_{\bS}^2\right)V=0
$$
Next, introduce $x\equiv\bS/S$ and set $\hat{V}(x,t)\equiv V/S$.
This reduces the dimension of the PDE by one
$$
  \left(\partial_t+\half\s^2(x-\phi)^2
  \partial_x^2\right)\hat{V}=0
$$
The resulting PDE is closely related to the one found by
Rogers and Shi~\cite{RogersShi95}. In fact, they can be
transformed into each other by a variable change $y=x-\phi$.
But our derivation of the PDE seems to be more natural:
working in a basis of tradables guides us in the right
direction. At this point, we will make the specific choice
of an exponential weight function
$$
  w(t) = \frac{e^{-\g(T-t)}}{T}
$$
Remember that if the interest rate is constant and equal to $r$,
then the choice $\g=r$ leads to an equally weighted average
in terms of the dollar price of the stock. We now find
$$
  \phi(t)=\frac{1-e^{-\g(T-t)}}{\g T}
$$
A change of variables is now in place
$$
\t\equiv T-t, \hspace{5mm}
z\equiv\frac{2e^{-\g\t}}{\s^2T(x-\phi)}
 =\frac{2Se^{-\g\t}}{\s^2TAP},
\hspace{5mm} s\equiv\s^2\t,
\hspace{5mm} \k\equiv\frac{\g}{\s^2}
$$
This transforms the PDE to
$$
\left(-\partial_s+\left((\k+1)z-\half z^2\right)\partial_z
+\half z^2\partial_z^2\right)\hat{V}=0
$$
A Laplace-transform with respect to $s$ yields
$$
\left(-\l+\left((\k+1)z-\half z^2\right)\partial_z
+\half z^2\partial_z^2\right)u=-f(z)
$$
where $u(z,\l)$ is the transformed function, and
$f(z)=\hat{V}(z,T)$ denotes the payoff at maturity. Next,
let us define
$$
u\equiv e^{\half z}z^{-\k-1}w
$$
Then $w$ satisfies
\begin{equation}
\left(\partial_z^2-\frac{1}{4}+\frac{(\k+1)}{z}
+\frac{(\frac{1}{4}-\m^2)}{z^2}\right)w=
-2e^{-\half z}z^{\k-1}f(z) \label{eq:whit}
\end{equation}
where we introduced
$$
\m\equiv\sqrt{\left(\k+\shalf\right)^2+2\l}
$$
The homogeneous part of Eq.~\ref{eq:whit} is Whittaker's
equation, and its solutions are the Whittaker functions
$M_{\k+1,\m}(z)$ and $W_{\k+1,\m}(z)$ (see appendix).
To solve Eq.~\ref{eq:whit} we will make use of a Green's
function approach. A Green's function with proper behaviour
at the boundaries is given by
$$
  G(x,y) = \frac{1}{Q}
    \bigg(M_{\k+1,\m}(x)W_{\k+1,\m}(y){\bf 1}_{x<y}+
    W_{\k+1,\m}(x)M_{\k+1,\m}(y){\bf 1}_{x>y}\bigg)
$$
where $Q$ is the Wronskian of the two solutions
$$
Q = W_{\k+1,\m}(z)\partial_z M_{\k+1,\m}(z)-
    M_{\k+1,\m}(z)\partial_z W_{\k+1,\m}(z)=
    \frac{\G(1+2\m)}{\G(-\half-\k+\m)}
$$
In terms of this, the solution can be written as
$$
  u(z,\l)=2e^{\half z}z^{-\k-1}\int_0^\infty G(x,z)
  e^{-\half x}{x}^{\k-1}f(x)dx
$$
The ASP payoff defined in Eq.~\ref{eq:payoff} corresponds
to the choice
$$
  f(z)=\left(\frac{2}{\s^2 T z}-k\right)^+=
  k\left(\frac{a}{z}-1\right)^+, \hspace{5mm}
  a\equiv\frac{2}{\s^2 kT}
$$
Inserting this in the integral, we find that the solution
falls apart in two ranges, $z<a$ and $z\geq a$
(or, equivalently, $e^{-\g\t}kS<AP$ and $e^{-\g\t}kS\geq AP$).
For details of this calculation we refer to the appendix.
In the former case, we find
$$
  \begin{aligned}
  u(z,\l)&=\frac{2ke^{\frac{z-a}{2}}z^{-\k-1}
  a^{\k}\G(-\half-\k+\m)}
  {\G(1+2\m)}W_{\k-1,\m}(a)M_{\k+1,\m}(z)\\
  &+\frac{2k\bigg(z\big((\k-\half)^2-\m^2\big)
  +a\big(z-\big((\k+\half)^2-\m^2\big)\big)\bigg)}
  {z\big((\k-\half)^2-\m^2\big)\big((\k+\half)^2-\m^2\big)}
  \end{aligned}
$$
The second term is exactly the Laplace transform of $x-k$.
In the latter case we find
$$
  u(z,\l)=\frac{2ke^{\frac{z-a}{2}}z^{-\k-1}
  a^{\k}\G(-\half-\k+\m)}
  {(-\half+\k+\m)(\half+\k+\m)\G(1+2\m)}
  M_{\k-1,\m}(a)W_{\k+1,\m}(z)
$$
Therefore, the solution can be written as
$$
  V_{\mbox{ASP}}(k,\g,t,T) = \left\{ \begin{array}{ll}
  kS(t)I_1+\bS(t)-kS(t) & e^{-\g\t}kS<AP \\
  kS(t)I_2 & e^{-\g\t}kS\geq AP \end{array} \right.
$$
where $I_1$ and $I_2$ are defined by inverse Laplace-transforms
$$
  I_1\equiv\frac{e^{\frac{z-a}{2}}z^{-\k-1}a^{\k}}{\p i}
  \int_{\rho-i\infty}^{\rho+i\infty}\frac{\G(-\half-\k+\m)
  W_{\k-1,\m}(a)M_{\k+1,\m}(z)e^{\l s}}{\G(1+2\m)}d\l
$$
$$
  I_2\equiv\frac{e^{\frac{z-a}{2}}z^{-\k-1}a^{\k}}{\pi i}
  \int_{\rho-i\infty}^{\rho+i\infty}\frac{\G(-\half-\k+\m)
  M_{\k-1,\m}(a)W_{\k+1,\m}(z)e^{\l s}}
  {(-\half+\k+\m)(\half+\k+\m)\G(1+2\m)}d\l
$$
where $\rho$ is an arbitrary constant chosen so that the contour
of integration lies to the right of all singularities in the
integrand. These integrals can be evaluated 
numerically~\cite{AbateWhitt95,Shaw98}. Note that the value of
an average strike call (ASC) follows simply from put-call parity,
that is, we use
$$
  \bigg(\bS(T)-kS(T)\bigg)^+ 
- \bigg(kS(T)-\bS(T)\bigg)^+ = \bS(T)-kS(T)
$$
and find
$$
  V_{\mbox{ASC}}(k,\g,t,T) = \left\{ \begin{array}{ll}
  kS(t)I_1 & e^{-\g\t}kS<AP \\
  kS(t)I_2-\bS(t)+kS(t) & e^{-\g\t}kS\geq AP \end{array} \right.
$$
The expression for the value of the ASP at $t=0$, i.e. its
unseasoned value, simplifies considerably. In this case
$z\rightarrow\infty$ and $s=\s^2 T=2/(ak)$, and we find
$$
V_{\mbox{ASP}}(k,\g,0,T)=
kS(0)\frac{e^{-\frac{a}{2}}a^{\k}}
{\pi i}\int_{\rho-i\infty}^{\rho+i\infty}
\frac{\G(-\half-\k+\m)
M_{\k-1,\m}(a)\exp(\frac{2\l}{ak})}
{(-\half+\k+\m)(\half+\k+\m)\G(1+2\m)}d\l
$$

\subsection{T-duality and average price options}
\label{sec:tdual2}

In this section we will use the T-duality, found in
section~\ref{sec:tdual}, to relate prices
of unseasoned average strike and average price options.
We will focus on the case where the weight function
is exponential. To indicate the weight function used
in the definition of $\bS$, we use a subscript $\g$,
i.e. we define
$$
  \bS_\g(t) = \frac{1}{T} \int_0^T e^{-\g(T-s)}Y_s(t)ds
$$
It is a straightforward calculation to see that the action
of the duality Eq.~\ref{eq:dual} (remark: we use a continuum
limit of this result) on the tradables $\bS_\g$, $S$ and $P$
is given by
$$ \boxed{ \begin{aligned}
  \bS_\g(T) &\leftrightarrow e^{-\g T}\bS_{-\g}(T) \\
  S(T) &\leftrightarrow \frac{S(0)}{P(0)}P(T)
\end{aligned} }
$$
Applying this to the payoff of an average price call gives
$$
  (\bS_\g(T)-KP(T))^+ \leftrightarrow
  e^{-\g T}\left(\bS_{-\g}(T)-\frac{KP(0)}
  {e^{-\g T}S(0)}S(T)\right)^+
$$
i.e., it transforms it into the payoff of an ASP.
Therefore we see that the unseasoned value of an APC
can be expressed as
$$
  V_{\mbox{APC}}(K,\g,0,T)=
  e^{-\g T}V_{\mbox{ASP}}
  \left(\frac{KP(0)}{e^{-\g T}S(0)},-\g,0,T\right)
$$
Similarly, we obtain the value of an unseasoned
average price put (APP)
$$
  V_{\mbox{APP}}(K,\g,0,T)=
  e^{-\g T}V_{\mbox{ASC}}
  \left(\frac{KP(0)}{e^{-\g T}S(0)},-\g,0,T\right)
$$
In this way, we actually reproduce the well known results
by Geman and Yor~\cite{GemanYor93}.

\subsection{On seasoned Asians}

It is a well-known fact that the price of a seasoned
average price option can be expressed in terms of the price
of an unseasoned average price option with a different strike.
Let us look at the mechanism behind this. We consider
an exponentially weighted, continuously sampled Asian with
a total lifetime of $M$, expiring at time $T$ (so it is
initiated at time $T-M<0$) and we are interested in its
price at $t=0$. As before, the payoff of such an option can
be expressed in terms of tradables $S$, $P$ and
$$
  \bS_{\g,M}(t) = \frac{1}{M}\int_{T-M}^T e^{-\g(T-s)}Y_s(t) ds
$$
where we explicitly show the longer sample period in the
definition by the subscript $M$. Now if $t\in [0,T]$ we can write
\begin{equation}
  \bS_{\g,M}(t) = \frac{T\bS_\g(t)}{M}+
  \frac{P(t)}{M}\int_{T-M}^0 e^{-\g(T-s)}\frac{S(s)}{P(s)} ds \equiv
  \frac{T \bS_\g(t) + AP(t)}{M} \label{eq:subst}
\end{equation}
where $A$ is proportional to the average over the time period
up to $t=0$. Substituting this in the payoff of an APC, we get
$$
  \bigg(\bS_{\g,M}(T)-KP(T)\bigg)^+ 
= \frac{T}{M}\bigg(\bS_\g(T)-\hat{K}P(T)\bigg)^+
$$
with
$$
  \hat{K} = \frac{MK-A}{T}
$$
This shows that the value of the {\it seasoned} APC that we
are considering can be expressed in terms of the value of an
{\it unseasoned} APC with a modified strike as
$$
  \frac{T}{M} V_{\mbox{APC}}(\hat{K},\g,0,T)
$$
Of course, the same trick also works for average price puts.
Note that it is possible for the strike $\hat{K}$ to become
negative. In that case, the option becomes trivial.
One might wonder what happens if we substitute Eq.~\ref{eq:subst}
in the payoff of an ASC. It turns out that in that case, things
do not combine in a nice way. Indeed, we find
\begin{equation}
  \bigg(kS(T)-\bS_{\g,M}(T)\bigg)^+ 
= \left(\frac{MkS(t)-T\bS_\g(T)-AP(T)}
  {M}\right)^+ \label{eq:sasc}
\end{equation}
Fortunately, we already have an expression for the value of
a seasoned ASC. So we can use the formula in reverse, to price
options with a payoff given by the RHS of Eq.~\ref{eq:sasc}.
This will turn out to be useful in the next section.

\section{Cash-dividend}

It well known in the literature how to price options
on a stock paying a known dividend yield. However, in many
cases it is more realistic to assume that the cash amount
of the dividend rather than the yield is known in advance.
This makes the pricing problem considerably harder.
In this section we show that the problem is equivalent
to the pricing of Asian options. In fact, we show that
their prices are connected by the put-call symmetries
of section~\ref{sec:putcall}. This allows us to use all
the techniques for the valuation of Asian options in the
context of options involving cash-dividend. The setting
is as follows. We assume that the stock follows geometric
Brownian motion between dividend payments, with fixed
volatility $\s$ (taking the bond as numeraire).
Dividends are paid at a set of discrete times
$\{t_i\}$, $i=1,2,3,\ldots$, $0<t_1<t_2<\cdots$ and are
expressed in units of the bond $\d(t_i)P$. Since we
will be interested in options with maturity $T$, we will
use a bond with this same maturity. We assume that the
interest rate is fixed and equal to $r$, so the bond
has dollar value $e^{-r(T-t)}$. By $S_i(t)$ we
mean the price of the stock between $t_i$ and $t_{i+1}$,
in other words
$$
  S(t) = S_i(t), \hspace{5mm} \mbox{for $t_i\leq t<t_{i+1}$}
$$
So at $t=t_1$ a portfolio consisting of 1 stock becomes
$$
  S_0 \rightarrow \d(t_1) P + S_1
$$
In order to avoid arbitrage, we assume that the
left-hand side equals the right-hand side at $t_1$.
This can be used to extend the definition of $S_0$
to all $t<t_2$ as follows
$$
  S(t) = S_1(t) = 
  \left(1-\frac{\d(t_1) P(t_1)}{S_0(t_1)}\right) S_0(t),
  \hspace{5mm} t_1\leq t < t_2
$$
Note that $S_0$, by construction, does {\it not} make a jump at
$t_1$. In fact, $S_0$ corresponds to the value of the self-financing
portfolio that one gets by directly reinvesting the cash-dividend
payment into the stock again. So $S_0$ is a tradable object.
We can repeat the process for the dividend payment at $t_2$
$$
  S_1 \rightarrow \d(t_2) P + S_2
$$
Again, $S_2$ can be expressed in terms of $S_0$, extending
the definition of $S_0$ to all $t<t_3$
$$
  S(t) = S_2(t) 
  = \left(1-\frac{\d(t_2) P(t_2)}{S_1(t_2)}\right) S_1(t) =
$$
$$
  \left(1-\frac{\d(t_1) P(t_1)}{S_0(t_1)}
         -\frac{\d(t_2) P(t_2)}{S_0(t_2)}\right) S_0(t),
  \hspace{5mm} t_2\leq t < t_3
$$
By repeating this process, we find that the value of a portfolio
$V$ which we get by starting with one stock at $t=0$ and holding it,
together with all its cumulative dividends up to time $t$
(note that this portfolio is also a tradable object, while the stock
by itself is not) is given by
$$
  V(t)=S(t)+\sum_{t_i\leq t} \d(t_i) P(t)=
  \left(1-\sum_{t_i\leq t}\frac{\d(t_i) P(t_i)}{S_0(t_i)}\right)S_0(t)
  +\sum_{t_i\leq t} \d(t_i) P(t)
$$
where $S_0$ just follows a lognormal price process.
Now if we consider a European option of the stock with
maturity $T$, we can define the cumulative dividends up
to maturity as follows
$$
  C(t) \equiv \sum_{t_i\leq T} \d(t_i) P(t)
$$
Using this we can write
$$
  V(t)=S_0(t)+C(t)-\bar{P}(t), \hspace{5mm} t\leq T
$$
where
$$
  \bar{P}(t) \equiv \sum_{t<t_i\leq T} \d(t_i) P(t)
   + \sum_{t_i\leq t} \frac{\d(t_i) P(t_i)}{S_0(t_i)} S_0(t)
$$
In terms of these new tradables, we can write
$$
  S(T) = S_0(T)-\bar{P}(T)
$$
Now the connection with Asian option becomes clear.
They can be transformed into each other by the exchange of
$S_0$ and $P$, i.e. by using put-call symmetry. In fact, we
can introduce tradable objects, similar to the $Y_s(t)$, as follows
$$
  X_s(t) = \left\{ \begin{array}{ll} P(t) & t\leq s \\
  \frac{P(s)}{S_0(s)} S_0(t) & t\geq s \end{array} \right.
$$
In terms of these, we can write
$$
  \bar{P}(t) = \sum_{t_i\leq T} \d(t_i) X_{t_i}(t)
$$
Again, for $t\in[0,T]$, we have
$$
  \begin{aligned}
  P(t) &= X_T(t) \\
  S_0(t) &= \frac{S(0)}{P(0)} X_0(t)
  \end{aligned}
$$
Therefore we see that the payoff of a plain vanilla option
on a stock paying cash dividends takes the general form
$$
  \left(\sum_i \d(t_i) X_{t_i}(T) \right)^+
$$
with certain weights $\d$. Taking $S_0(t)$ as numeraire, we
see that the tradables $X_s(t)$ satisfy
$$
  dX_s(t) = {\bf 1}_{t<s}\s X_s(t) dW(t) + \cdots
$$
and we can use the integral approach described in
section~\ref{sec:mulint} to price the option. Alternatively
we can use a PDE approach, i.e. we generalize the definition
of $\bar{P}$ as follows, cf. the steps in \ref{sec:pdeapp},
$$
  \bar{P}(t)= \int_0^T \d(s) X_s(t) ds
  = \phi(t)P(t)+A(t)S_0(t) 
$$
A PDE which describes the price process of an option
depending on $S_0$, $P$ and $\bar{P}$ (this class includes
plain vanilla options on stocks paying cash-dividends)
can now easily be derived (just take the stock as numeraire)
$$
  \left(\partial_t+\half\s^2P^2
  (\partial_P+\phi\partial_{\bar{P}})^2\right)V=0
$$
It is instructive to consider a call option on a stock
which pays a continuous stream of cash dividends, with
exponential weights. One can use this as an approximation
to the value of an option where the underlying stock
pays a long stream of discrete cash dividends.
So let us define
$$
  \bar{P}_\g(t) \equiv \frac{1}{T} \int_0^T e^{-\g(T-s)}X_s(t) ds
$$
The natural choice is $\g=-r$, which corresponds to a constant
dividend stream in terms of dollars. The payoff of a plain vanilla
call becomes
$$
  \bigg(S(T)-KP(T)\bigg)^+ \rightarrow 
  \bigg(S_0-\d T\bar{P}_{-r}(T)-KP(T)\bigg)^+
$$
where $\d$ parametrizes the dividend stream. We will from
now on omit the subscript from the $S_0$. By exchanging $S$
and $P$, exploiting put-call symmetry, we get
$$
  \bigg(P(T)-\d T\bS_{-r}(T)-KS(T)\bigg)^+
$$
which is the payoff of some Asian option. In fact, by using
T-duality, the payoff can be related to one that corresponds
to a seasoned average strike call (see Eq.~\ref{eq:sasc})
$$
  \left( \frac{P(0)}{S(0)}S(T)-\d T e^{r T}\bS_r(T)
        -\frac{S(0)}{P(0)}KP(T) \right)^+
$$
and we can use the analytical results that we derived for
this type of option to write down the price of this instrument.

\section{Conclusion and outlook}

In this article we have shown the power of symmetries to derive
prices of complex exotic options. We focused on arithmetic
average options in a Black-Scholes setting. By choosing an appropriate
basis of tradables, i.e. self-financing portfolios, it becomes a
straightforward matter to write down the governing PDE for the option
price. We then proceed to derive the Laplace-transformed price of a
European average strike option. This result extends the result
of Geman and Yor~\cite{GemanYor93} for the average price option.
Next we show the power of the underlying symmetry, by showing the
equivalence of the unseasoned arithmetic average strike and price
options after a suitable transformation of parameters. Seasoned
options can be treated in a similar way. Finally we exploit the
symmetry in the problem to show that vanilla options on stocks paying
cash dividends are equivalent, after suitable transformations, to
arithmetic Asian options, thus providing a method to price these type
of options.

\vspace{1\baselineskip}\noindent
Let us remark that the present discussion carries over without too
much changes to the case of basket options and swaptions. We will
discuss this in future work. Also we did not discuss the case of an
arithmetic Asian option with early-exercise features. This is however
simple to implement and we will come back to this in a future work.

% \section{Semi-static hedging of discretely sampled Asians}

% If we perform a change of variables in Eq.~\ref{eq:funpde},
% eliminating $\bar{S}$ in favor of $A$, we find
% %
% $$
% \left(\partial_t+w\frac{S}{P}\partial_A+
% \half\s^2S^2\partial^2_S\right)V=0
% $$
% %
% From this PDE we see that in between sample dates, where $w=0$,
% the price of a discretely sampled Asian option evolves just like
% a plain vanilla. This allows us to set up a semi-static hedging
% strategy as follows. Suppose we are at $t_m\leq t<t_{m+1}$,
% where $t_i$ are sample dates. With the information we have at that
% time, we can calculate the value of the option at time $t_{m+1}$
% as a function of $S$ and $P$ alone:
% %
% $$
%   f(S,P) = V(S,P,A(S,P),t_{m+1})
% $$
% %
% where $A(S,P)$ is the running average at time $t_{m+1}$, which
% includes the sample at that time. So between $t_m$ and
% $t_{m+1}$, the Asian option equals a European option with payoff
% $f(S,P)$ at $t_{m+1}$. Such an option can be hedged statically
% by the methods of Carr and Chou, using a portfolio consisting
% of plain vanilla puts and calls. Of course, at $t=t_{m+1}$
% we have to update the position, calculating a payoff for $t_{m+2}$.
% Continuing in this way, we find a semi-static hedge for the Asian.

\appendix
\section{Whittaker functions}

In this appendix we enumerate some useful properties
of Whittaker functions. More information can be found in
e.g.~~\cite{AbramowitzStegun64,PBM86}. The Whittaker functions
$M_{\k,\m}(z)$ and $W_{\k,\m}(z)$ are solutions to Whittaker's PDE
$$
  \left(\partial_z^2-\frac{1}{4}+\frac{\k}{z}
  +\frac{(\frac{1}{4}-\m^2)}{z^2}\right)f=0
$$
These functions are defined as
$$
M_{\k,\m}(z)=e^{-\half z}z^{\m+\half}\kummer(\shalf+\m-\k,1+2\m,z)
$$
where the confluent hypergeometric function is given by
$$
\kummer(a,b,z)=\frac{\G(b)}{\G(a)}
\sum_{n=0}^{\infty}\frac{\G(a+n)z^n}{\G(b+n)n!}
$$
and
$$
W_{\k,\m}(z)=e^{-\half z}z^{\m+\half}\Psi(\shalf+\m-\k,1+2\m,z)
$$
where $\Psi$ is the Tricomi function
$$
\Psi(a,b,z) = \frac{1}{\G(a)}\int_0^{\infty}
e^{-zt}t^{a-1}(1+t)^{b-a-1}dt
$$
There are the following interesting relations
$$
W_{\k,\m}(z)=\frac{\G(-2\m)}{\G(\half-\m-\k)}
M_{\k,\m}(z)+\frac{\G( 2\m)}{\G(\half+\m-\k)}
M_{\k,-\m}(z)
$$
$$
z^{-\m-\half}M_{\k,\m}(z)=
(-z)^{-\m-\half}M_{-\k,\m}(-z)
$$
To evaluate the price of the ASP we made use of
the following definite integrals
$$
\begin{aligned}
  \int e^{-\half z}z^{\k-1}W_{\k+1,\m}(z)dz&=
  -e^{-\half z}z^{\k}W_{\k,\m}(z)
\\
  \int e^{-\half z}z^{\k-2}W_{\k+1,\m}(z)dz&=
  e^{-\half z}z^{\k-1}(W_{\k-1,\m}(z)-W_{\k,\m}(z))
\\
  \int e^{-\half z}z^{\k-1}M_{\k+1,\m}(z)dz&=
  \frac{e^{-\half z}z^{\k}}{\G(\frac{3}{2}+\k+\m)}
  \G(\shalf+\k+\m) M_{\k,\m}(z)
\\
  \int e^{-\half z}z^{\k-2}M_{\k+1,\m}(z)dz&=
  \frac{e^{-\half z}z^{\k-1}}{\G(\frac{3}{2}+\k+\m)} \times
\\
  \times (\G(\shalf+\k+\m) &M_{\k,\m}(z)+
   \G(-\shalf+\k+\m) M_{\k-1,\m}(z))
\end{aligned}
$$
To calculate its unseasoned value, we recall the
asymptotic behaviour of $W_{\k,\m}(z)$
$$
W_{\k,\m}(z) \sim e^{-\half z}z^{\k}, \hspace{5mm}
z\rightarrow\infty
$$

\end{document}